# Fermi surface nesting induced strong pairing in iron-based superconductors

K. Terashima[a,1], Y. Sekiba[b], J. H. Bowen[c], K. Nakayama[b], T. Kawahara[b], T. Sato[b,d],

P. Richard[e], Y.-M. Xu[f], L. J. Li[g], G. H. Cao[g], Z.-A. Xu[g], H. Ding[c],

and T. Takahashi[b,e]

[a]UVSOR Facility, Institute for Molecular Science, Okazaki 444-8585, Japan

[b]Department of Physics, Tohoku University, Sendai 980-8578, Japan

[c]Beijing National Laboratory for Condensed Matter Physics, and Institute of Physics, Chinese Academy of Sciences, Beijing 100190, China

[d]TRIP, Japan Science and Technology Agency (JST), Kawaguchi 332-0012, Japan

[e]WPI Research Center, Advanced Institute for Materials Research, Tohoku University, Sendai 980-8577, Japan

[f]Department of Physics, Boston College, Chestnut Hill, MA 02467, USA

[g]Department of Physics, Zhejiang University, Hangzhou 310027, China

The discovery of high-temperature superconductivity in iron pnictides raised the possibility of an unconventional superconducting mechanism in multiband materials. The observation of Fermi-surface(FS)-dependent nodeless superconducting gaps suggested that inter-FS interactions may play a crucial role in superconducting pairing. In the optimally hole-doped $Ba_{0.6}K_{0.4}Fe_2As_2$, the pairing strength is enhanced simultaneously ($2\Delta/T_c \sim 7$) on the nearly nested FS pockets, *i.e.* the inner holelike ($\alpha$) FS and the two hybridized electronlike FSs, while the pairing remains weak ($2\Delta/T_c \sim 3.6$) in the poorly-nested outer hole-like ($\beta$) FS. Here we report that in the electron-doped $BaFe_{1.85}Co_{0.15}As_2$ the FS nesting condition switches from the $\alpha$ to the $\beta$ FS due to the opposite size changes for hole- and electron-like FSs upon electron doping. The strong pairing strength ($2\Delta/T_c \sim 6$) is also found to switch to the nested $\beta$ FS, indicating an intimate connection between FS nesting and superconducting pairing, and strongly supporting the inter-FS pairing mechanism in the iron-based superconductors.

In charge doped superconductors, such as copper oxides (cuprates), electron or hole doping may influence the superconducting (SC) properties differently (1, 2). As an example, angle-resolved photoemission spectroscopy (3) (ARPES) and Raman scattering (4) revealed a non-monotonic behaviour in the SC gap function of the electron-doped cuprates which is different from the simple $d_{x^2-y^2}$-wave function observed in the hole-doped cuprates (5). On the other hand, in the new Fe-based superconductors (6-9), no direct comparison of the SC order parameter has been made between hole- and electron-doped systems. ARPES studies on hole-doped $Ba_{1-x}K_xFe_2As_2$ have observed isotropic gaps that have different values on different Fermi surfaces (FSs) with strong pairing occurring on the nearly nested FS pockets (10-13). Thus, it is particularly important to conduct a comparison of the SC gaps and their FS dependence of an electron-doped pnictide. We have chosen $BaFe_{1.85}Co_{0.15}As_2$, which is optimally electron doped (14) with the same crystal structure as the $Ba_{1-x}K_xFe_2As_2$ system (9).

**Results**

Figures 1*A* and *B* show ARPES intensity plots of $BaFe_{1.85}Co_{0.15}As_2$ ($T_c$ = 25.5 K) as a function of binding energy and momentum (*k*) along two high symmetry lines in the Brillouin zone (BZ). We observe a holelike dispersion centred at the Γ point and two electronlike FSs near the M point. Even though a reasonable agreement is found between experiment and renormalized band calculations (15), some experimental features such as the energy position of the 0.2 eV band at the Γ point and the bottom of the electron band at the M point,

are not well reproduced by band calculations. This suggests a possible orbital and $k$-dependence of the mass-renormalization factor. Figure 1$C$ shows the ARPES intensity at the Fermi level ($E_F$) plotted as a function of the in-plane wave vector. A circular and an elongated intensity pattern centred at the Γ and M points are clearly visible, and they are attributed to the hole- and electron-like bands in Figs. 1$A$ and $B$. In Figures 1$D$ and $E$, we show ARPES intensity plots near $E_F$ and the corresponding energy distribution curves (EDCs) measured at three representative cuts indicated in Fig. 1$C$. In cut 1 of Fig. 1$E$, two holelike bands are clearly visible. The inner band, which is assigned to the likely degenerate α band (10, 16), sinks significantly (~30 meV) and does not create a small FS pocket as observed in the hole-doped samples (10), confirming the electron doping by the Co substitution. The outer band (the β band) crosses $E_F$ creating a hole pocket at the Γ point. We also notice that the number of holelike bands crossing $E_F$ in the present ARPES experiment is different from the band calculations (one and three, respectively). Along cut 3 which is close to M, we can distinguish an electronlike band crossing $E_F$ with a bottom at ~40 meV creating an elongated FS pocket shown in Fig. 1$C$. A closer look at the second derivative of momentum distribution curves (MDCs) in Fig. 1$F$ reveals the presence of an additional band whose dispersion near $E_F$ is nearly parallel to the main band. This suggests that the electronlike FSs consist of the inner (γ) and outer (δ) pockets resulting from the hybridization of two ellipsoidal pockets elongated along the $k_x$ and $k_y$ directions (see Fig. 1$C$), as in the case of hole-doped $Ba_{0.6}K_{0.4}Fe_2As_2$ (16). We remark here that the topology of the FS near the M point in hole-doped $Ba_{1-x}K_xFe_2As_2$ is controversial among several

ARPES groups. This controversy is closely related to the presence of a holelike band around the M point that intersects the electronlike bands in the vicinity of $E_F$, producing a complicated intensity distribution pattern around $E_F$. It is thus important to follow carefully the band dispersion with high energy and momentum resolutions when determining the FS topology (10-11). As clearly seen in Fig. 1C, the β hole pocket is nearly nested with the electron pockets, in sharp contrast to the observation of a good FS nesting between the α hole pocket and the electron pockets in the hole-doped $Ba_{0.6}K_{0.4}Fe_2As_2$.

We now illustrate how the SC gap evolves below $T_c$ on each FS. Figure 2A displays the temperature ($T$) dependence of the EDCs at a Fermi vector ($k_F$) on the β hole pocket measured across $T_c$. At 8 K, the midpoint of the leading edge below $T_c$ is shifted toward higher binding energy by ~4 meV with a pile-up of spectral weight at ~7 meV, indicating the opening of a SC gap. A similar gap value has been observed in the density of states (DOS) by scanning tunnelling spectroscopy on a same type of sample (17). We have eliminated the effect of the Fermi-Dirac distribution function by symmetrizing the EDCs at each temperature (18) (Fig. 2B). To cancel out the influence of the V-shaped spectral DOS, which originates from the tail of the α band, we divided each symmetrized spectrum by the 33-K normal state spectrum (see top of Fig. 2B). As one easily notices, a sharp coherence peak emerges below $T_c$ at 7 meV. Interestingly, the spectral weight at $E_F$ displays a small depression even at $T = 27$ K, indicating a possible weak pseudogap just above $T_c$. We have estimated the size of the SC gaps and plotted them in Fig. 2C. It is found that a simple Bardeen-Cooper-Schrieffer mean-field $T$-dependence with $\Delta(0) = 7$ meV reproduces satisfactorily the extracted gap amplitudes. In Figures 2*D* and *E*, we plot the *T*-dependence of raw and symmetrized EDCs measured on the ellipsoidal electron pocket. Although the overall *T*-dependence of the leading-edge shift and the spectral weight suppression below $T_c$ on the electron pockets are qualitatively similar to those of the β FS, there are some differences: a weaker coherence-peak weight accompanied with a reduction of spectral weight at $E_F$, and a smaller SC-gap size (4.5 *vs*. 6.7 meV, see Fig. 2*F*) are observed. These results demonstrate the FS-sheet dependence of the SC gap. We also noticed that the weight of the coherence peak is much weaker than that of hole-doped $Ba_{0.6}K_{0.4}Fe_2As_2$ (10), possibly owing to a lower superfluid density due to a lower $T_c$ value (25.5 *vs*. 37 K) and disorder scattering induced by in-plane Co substitution.

Next we turn our attention to the *k*-dependence of the SC gaps. Figures 3*A* and *B* show the symmetrized EDCs measured at 8 K at various $k_F$ points on (*A*) the β hole pocket and (*B*) the ellipsoidal electron pocket. As shown in Fig. 3*D*, the SC gap of each FS is nearly constant, indicating an isotropic *s*-wave nature. On the other hand, the average gap values of the hole and electron pockets are different (6.6 and 5.0 meV, respectively), establishing unambiguously the FS-dependent nature of the SC gap.

**Discussion**

A direct comparison of the ARPES data on the electron- and hole-doped SC pnictides allows us to conclude that: (i) the SC gap opens on multiple FSs centred at the Γ and M points, (ii) the SC gap is nodeless and exhibits nearly isotropic behaviour on each FS, and (iii) the pairing strength, as reflected by the ratio of $2\Delta/k_BT_c$, is related to the FS nesting condition between the electron and hole pockets. In hole-doped $Ba_{0.6}K_{0.4}Fe_2As_2$ (10, 11), the interband scattering via the wavevector $Q\sim(\pi,0)$ (as defined in the unreconstructed BZ) has been suggested to enhance the pairing amplitude of the α and γ(δ) FSs (19-22), resulting in large $2\Delta/k_BT_c$ values of 7.2-7.7, while the poorly-nested β FS has a value of 3.6, close to the weak-coupling regime. Remarkably, in electron-doped $BaFe_{1.85}Co_{0.15}As_2$, the β (but not α) is connected to the γ(δ) FSs by the same $Q\sim(\pi,0)$ and possess strong-coupling $2\Delta/k_BT_c$ values of 5.9, suggesting an enhancement of the pairing amplitude due to inter-pocket scattering on the nearly nested FSs. The observation that the pairing strength in the β band increases from 3.6 in the optimally hole-doped sample to 5.9 in the optimally electron-doped sample strongly suggests that the SC coupling strength is more related to the nesting condition among the FSs than to the orbital characters themselves. All these experimental observations suggest that the inter-pocket scattering and Fermi surface nesting are critical aspects of the pairing mechanism of the pnictides.

At this point, a few essential issues regarding the nature of the unconventional pairing mechanism need to be addressed. The first one is the cause of a small difference in the SC gap size of the observed FSs in the electron-doped system. This may be related to the difference in the partial DOS between electron and hole pockets. The existence of two electron pockets would give rise to a larger DOS at $E_F$ in the electron pockets than in the hole pocket. This leads to a relative enhancement of the pairing amplitude in the hole pocket due to stronger $Q\sim(\pi,0)$ scattering from the electron pockets (23). The second issue concerns the smaller $2\Delta(0)/k_BT_c$ values obtained on the well nested FSs in the electron-doped $BaFe_{1.85}Co_{0.15}As_2$ (4.5-5.9) as compared to the hole-doped $Ba_{0.6}K_{0.4}Fe_2As_2$ (7.2-7.7 (10)). This might be linked to the larger pairing-breaking disorder scattering caused by the in-plane Co substitution as compared to the off-plane K substitution in $Ba_{0.6}K_{0.4}Fe_2As_2$. The last issue, which might be the most unusual, is the lack of anisotropy in the gap function on a given FS, in contrast to a theoretical prediction of anisotropic extended *s*-wave gaps based on renormalization group calculation (24). Strictly speaking, the interband scattering condition *via* $Q = (\pi,0)$ is not perfect in either the hole or the electron-doped systems since the shape of the hole and electron pockets does not match completely (*e.g.* see Fig. 1*C*). One might therefore expect a large anisotropy in the SC gap size near the M point, while the observed SC gap size is merely dependent on the FS sheet. This implies that there might be a novel mechanism that keeps the intraband SC gap constant. Our observation of isotropic SC gaps that depend on FS nesting conditions is an important step toward a full understanding of high temperature superconductivity in iron pnictides.

**Materials and Methods**

**Samples preparation.** High-quality single crystals of $BaFe_{1.85}Co_{0.15}As_2$ were grown by the self flux method, the same growth method as for $BaFe_{2-x}Ni_xAs_2$ (25). From electrical resistivity measurements, $T_c$ of the sample has been estimated to be $T_c^{mid}$ = 25.5 K. The starting material (nominal composition) was $BaFe_{1.8}Co_{0.2}As_2$, while the actual Co content was determined by energy-dispersive X-ray spectroscopy. The low-energy electron diffraction (LEED) patterns indicate that the sample surface has a 4-fold symmetry with no signature of surface reconstruction.

**ARPES experiments.** ARPES measurements were performed using a VG-SCIENTA SES2002 spectrometer with a high-flux discharge lamp and a toroidal grating monochromator. We used the He I$\alpha$ resonance line (21.218 eV) with a partially linear polarization parallel to the $\Gamma M$ ($\Gamma X$) direction for measurements of the $\Gamma M$ ($\Gamma X$) cut, as shown in Fig. 1*C* (orange arrows). The energy resolution was set at 4 and 15 meV for the measurement of the SC gap and the valence band dispersion, respectively. The angular resolution was set to 0.2°. Clean surfaces for ARPES measurements were obtained by *in-situ* cleaving of crystals at 8 K in a working vacuum better than $5 \times 10^{-11}$ Torr, and then measured immediately without annealing. Mirror-like sample surfaces were found to be stable without obvious degradation for the measurement period of 3 days. The Fermi level of the samples was referenced to that of a gold film evaporated onto the sample holder. We observed that the SC gap closes around the bulk $T_c$ (see Fig. 2), which suggests that the present ARPES results reflect the bulk properties. We found that the periodicity of the experimental band dispersions matches well with that of the bulk crystal, suggesting that no surface relaxation or reconstruction takes place, as also supported by the measured LEED patterns. We have confirmed the reproducibility of the ARPES data on more than five sample surfaces and through thermal cycles across $T_c$. To extract the carrier number from the ARPES measurement, we have estimated the FS volume of the β pocket and the ellipsoidal electron pocket to be 1.6% and 3.2% of the unfolded first BZ respectively. The deduced total carrier number of 0.05 electrons/Fe is close to but slightly lower than the expected value of 0.075 electrons/Fe in $BaFe_{1.85}Co_{0.15}As_2$. The small difference may suggest a possible deviation of the Co valency from 3+ and the finite three dimensionality of the band structure (19, 26-28).

**Acknowledgments**

We thank X. Dai, Z. Fang, and Z. Wang for providing their band calculation results and valuable discussions. K. T. and K. N. thank JSPS for financial support. This work was supported by grants from JSPS, TRIP-JST, CREST-JST, MEXT of Japan, the Chinese Academy of Sciences, NSF, Ministry of Science and Technology of China, and NSF of US.

**References**

1. Damascelli A, Hussain Z, Shen ZX (2003) Angle-resolved photoemission studies of the cuprate superconductors. *Rev Mod Phys* 75:473-541.

2. Campuzano JC, Norman MR, Randeria M (2004) in *The Physics of Superconductors* eds Bennemann KH, Ketterson JB (Springer, Berlin), Vol. II pp 167-273.

3. Matsui H, *et al*. (2005) Direct observation of a nonmonotonic $d_{x^2-y^2}$-wave superconducting gap in the electron-doped high-$T_c$ superconductor $Pr_{0.89}LaCe_{0.11}CuO_4$. *Phys Rev Lett* 95:017003.

4. Blumberg G, *et al*. (2002) Nonmonotonic $d_{x^2-y^2}$ superconducting order parameter in $Nd_{2-x}Ce_xCuO_4$. *Phys Rev Lett* 88:107002.

5. Ding H, *et al*. (1996) Angle-resolved photoemission spectroscopy study of the superconducting gap anisotropy in $Bi_2Sr_2CaCu_2O_{8+x}$. *Phys Rev B* 54:R9678.

6. Kamihara Y, Watanabe T, Hirano M, Hosono H (2008) Iron-based layered supeconductor $La[O_{1-x}F_x]FeAs$ (x = 0.05-0.12) with $T_c$ = 26 K. *J Am Chem Soc* 130:3296-3297.

7. Takahashi H, *et al*. (2008) Superconductivity at 43 K in an iron-based layered compound $LaO_{1-x}F_xFeAs$. *Nature* 453:376-378.

8. Chen XH, *et al*. (2008) Superconductivity at 43 K in $SmFeAsO_{1-x}F_x$. *Nature* 453:761-762.

9. Rotter M, Tegel M, Johrendt D (2008) Superconductivity at 38 K in the iron arsenide $(Ba_{1-x}K_x)Fe_2As_2$. *Phys Rev Lett* 101:107006.

10. Ding H, *et al*. (2008) Observation of Fermi-surface-dependent nodeless superconducting gaps in $Ba_{0.6}K_{0.4}Fe_2As_2$. *Europhys Lett* 83:47001.

11. Nakayama K, *et al*. (2008) Superconducting-gap symmetry of $Ba_{0.6}K_{0.4}Fe_2As_2$ studied by angle-resolved photoemission spectroscopy. arXiv:0812.0663v1 [cond-mat.supr-con].

12. Liu C, *et al*. (2008) K-doping dependence of the Fermi surface of the iron-arsenic $Ba_{1-x}K_xFe_2As_2$ superconductor using angle-resolved photoemission spectroscopy. *Phys Rev Lett* 101:177005.

13. Zhao L, *et al*. (2008) Unusual superconducting gap in $(Ba,K)Fe_2As_2$ superconductor. *Chin Phys Lett* 25:4402-4405.

14. Sefat AS, *et al*. (2008) Superconductivity at 22 K in Co-doped $BaFe_2As_2$ crystals. *Phys Rev Lett* 101:117004.

15. Xu G, Zhang H, Dai X, Fang Z (2008) Electron-hole asymmetry and quantum critical point in hole-doped $BaFe_2As_2$. *Europhys Lett* 84:67015.

16. Ding H, *et al*. (2008) Electronic structure of optimally doped pnictide $Ba_{0.6}K_{0.4}Fe_2As_2$: a comprehensive ARPES investigation. arXiv:0812.0534v1 [cond-mat.supr-con].

17. Yin Y, *et al*. (2009) Scanning tunneling spectroscopy and vortex imaging in the iron pnictide superconductor $BaFe_{1.8}Co_{0.2}As_2$. *Phys Rev Lett* 102:097002.

18. Norman MR, *et al*. (1998) Destruction of the Fermi surface in underdoped high $T_c$ superconductors. *Nature* 392:157-160.

19. Mazin II, Singh DJ, Johannes MD, Du MH (2008) Unconventional superconductivity with a sign reversal in the order parameter of $LaFeAsO_{1-x}F_x$. *Phys Rev Lett* 101:057003.

20. Kuroki K, *et al*. (2008) Unconventional pairing originating from the disconnected Fermi surfaces of superconducting $LaFeAsO_{1-x}F_x$. *Phys Rev Lett*

101:087004.

21. Wang F, *et al*. (2009) A functional renormalization group study of the pairing symmetry and pairing mechanism of the FeAs based high temperature superconductors. *Phys Rev Lett* 102:047005.

22. Seo K, Bernevig BA, Hu J (2008) Pairing symmetry in a two-orbital exchange coupling model of oxypnictides. *Phys Rev Lett* 101:206404.

23. Dolgov OV, Mazin II, Parker D, Golubov AA (2009) Interband superconductivity: contrasts between BCS and Eliashberg theory. *Phys Rev B* 79:060502.

24. Wang F, Zhai H, Lee DH (2009) Antiferromagnetic Correlation and the Pairing Mechanism of the Cuprates and Iron Pnictides : A view from the functional renormalization group studies. *Europhys Lett* 85:37005.

25. Li LJ, *et al*. (2009) Superconductivity induced by Ni doping in $BaFe_2As_2$. *New J Phys* 11:025008.

26. Singh DJ (2008) Electronic structure and doping in $BaFe_2As_2$ and LiFeAs: Density functional calculations. *Phys Rev B* 78:094511.

27. Nekrasov IA, Pchelkina ZV, Sadovskii MV (2008) Electronic structure of prototype $AFe_2As_2$ and ReOFeAs high-temperature superconductors: a comparison. *JETP Lett* 88:144-149.

28. Gordon RT, *et al*. (2008) Unconventional London penetration depth in $Ba(Fe_{0.93}Co_{0.07})_2As_2$ single crystals. arXiv:0810.2295v1 [cond-mat.supr-con].

**Figure legends**

FIG. 1(COLOUR)

Fermi surface and band structure of electron-doped $BaFe_{1.85}Co_{0.15}As_2$.

(*A, B*) ARPES intensity plots of $BaFe_{1.85}Co_{0.15}As_2$ ($T_c$ = 25.5 K) as a function of wave vector and binding energy measured at 8 K along (*A*) the $\Gamma X$ and (*B*) the $\Gamma M$ lines with the He I$\alpha$ ($h\nu$ = 21.218 eV) resonance line, together with the band dispersion from the first-principle calculations for $k_z$ = 0 and $\pi$ (blue and red curves, respectively). Calculated bands for $BaFe_2As_2$ (15) were shifted downward by 40 meV and then renormalized by the factor of 2. Green broken lines in *A* and *B* denote the expected $E_F$ positions of $BaFe_2As_2$ and $Ba_{0.6}K_{0.4}Fe_2As_2$. (*C*) FS contour determined by plotting the ARPES spectral intensity integrated within ±5 meV with respect to $E_F$. Black filled circles in *C* show the $k_F$ positions determined by tracing the experimental band dispersion, while blue open circles are symmetrized $k_F$ points obtained by assuming a 4-fold symmetry with respect to the $\Gamma$ and M point, respectively. The symmetrized $k_F$ points (blue open circles) coincide well with the original ones (black filled circles), confirming the validity of this symmetry operation. Orange arrows show the polarization vector of the incident light for each cut. (*D*) ARPES spectral intensity at 8 K as a function of wave vector and binding energy, and (*E*) corresponding EDCs measured along three representative cuts 1-3 shown in *C*. Circles in *E* traces the energy dispersion of the $\alpha$ and $\beta$ bands. (*F*) Second derivative plot of MDCs along cut 3 to highlight the presence of another weaker electronlike band ($\delta$).

FIG 2 (COLOUR)

Temperature dependence of the superconducting gap.

(A) T-dependence of EDC measured at a $k_F$ point on the β FS (red dot in inset). (B) Symmetrized EDCs and the same but divided by the spectra at $T$ = 33 K. Dashed line denotes the position of SC coherence peak. (C) T-dependence of the SC gap size. Solid line is the BCS mean-field gap with $T_c$ = 25.5 K and $\Delta(0)$ = 7 meV. (D-F) same as A-C but measured on the $k_F$ point of the ellipsoidal electron pocket. Dashed line in F is the same as the solid line in C.

FIG. 3(COLOUR)

Momentum dependence of the superconducting gap.

(A, B) Symmetrized EDCs at 8 K measured at various $k_F$ points on the β and electronlike FS, labelled by respective coloured symbols correspondingly. (C) Extracted FS from the ARPES measurements together with the definition of FS angle ($\theta$). (D) SC gap values at 8 K as a function of $\theta$ extracted from the EDCs shown on the polar plot, for the β and electronlike FSs (red and blue dots, respectively). Dashed circles represent the averaged gap value.

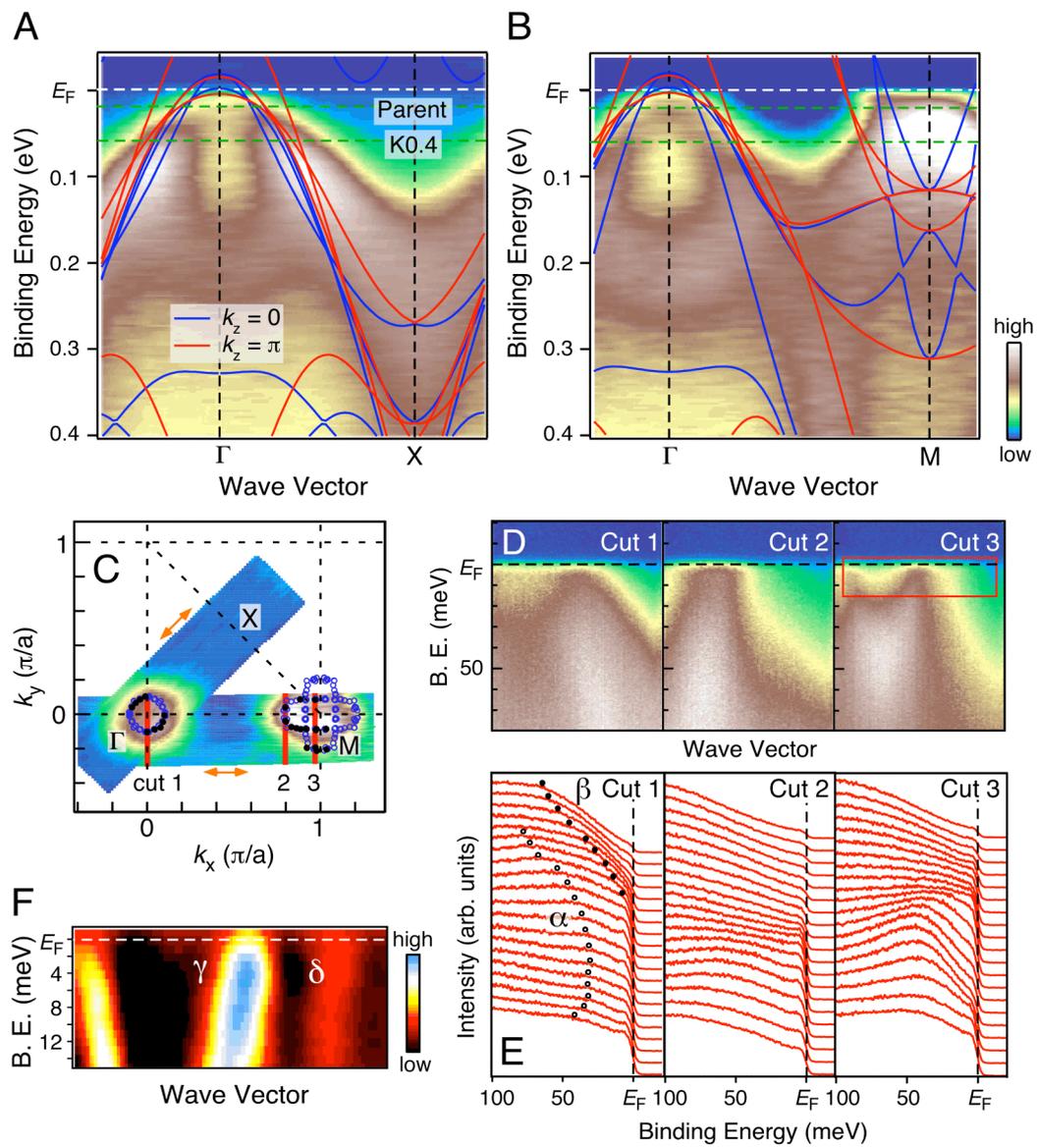

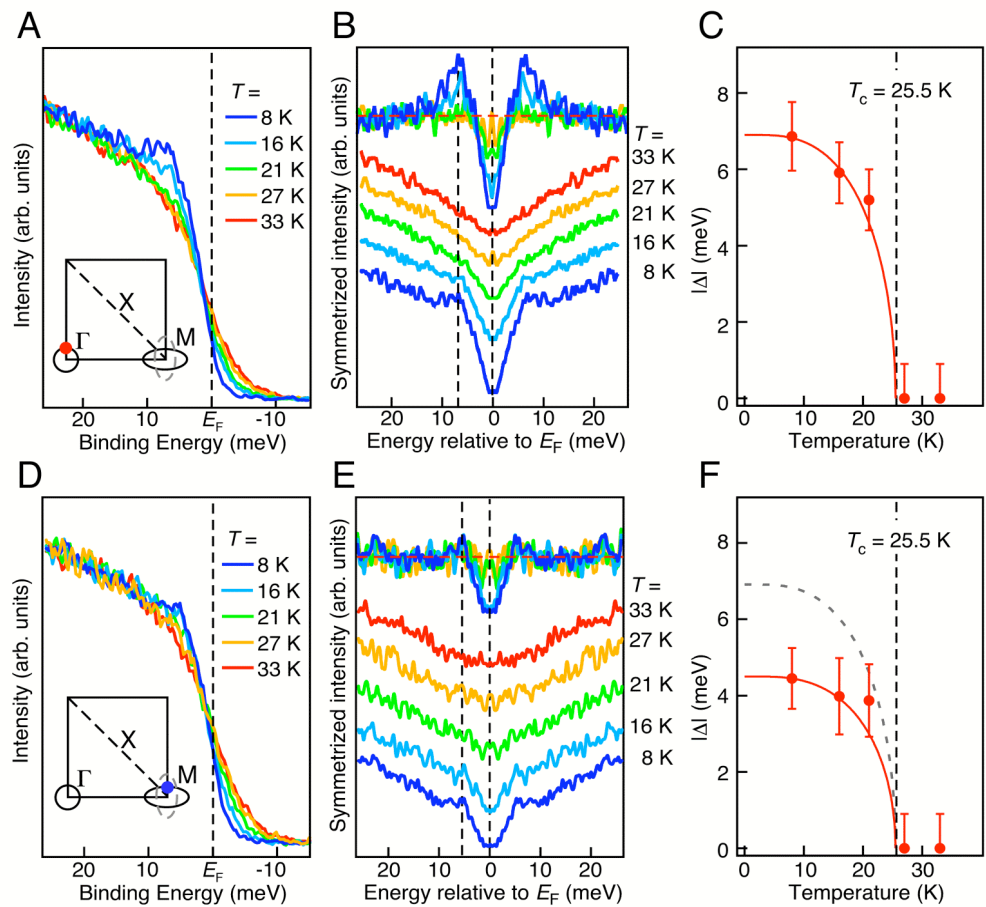

Figure 2

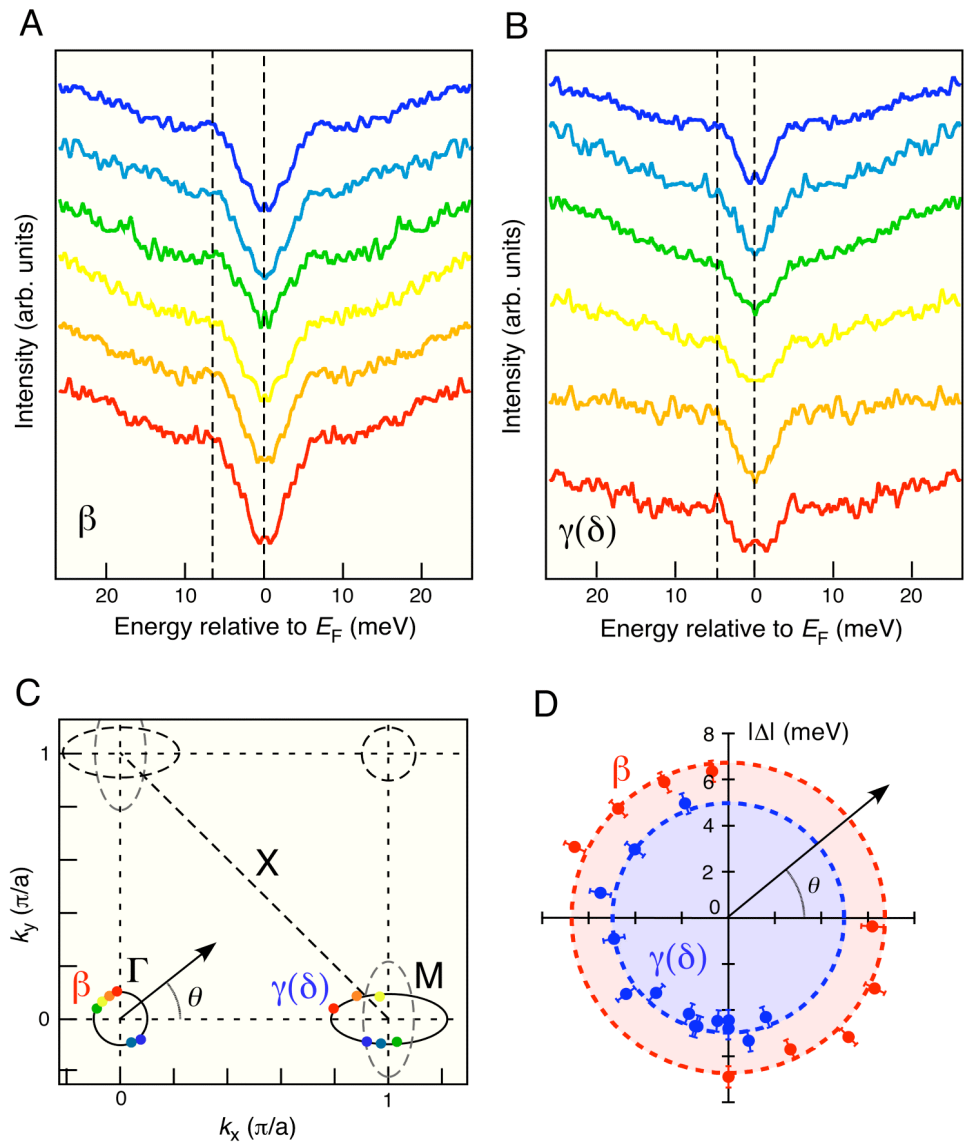

Figure 3